\documentclass[12pt,aps,prb,preprint,showpacs,showkeys,groupedaddress]{revtex4}
\usepackage{graphicx}

\begin{document}


\title{Thump, ring: the sound of a bouncing ball}


\author{J. I. Katz}
\email[]{katz@wuphys.wustl.edu}
\affiliation{Department of Physics and McDonnell Center for the Space
Sciences, Washington University, St. Louis, Mo. 63130}


\date{\today}

\begin{abstract}
A basketball bounced on a stiff surface produces a characteristic loud
thump, followed by high-pitched ringing.  Describing the ball as an
inextensible but flexible membrane containing compressed air, I formulate
an approximate theory of the generation of these sounds and predict their
amplitudes and waveforms.
\end{abstract}


\maketitle


\section{Introduction}

A basketball bounced on a stiff surface, such as a thick concrete slab,
emits a loud characteristic ``thump'', followed by a high-pitched ringing.  
The sound is very different if the ball is bounced on a more resilient or
softer surface.  These characteristic sounds are not heard when a solid ball
is bounced, and the bounce of a soft rubber ball of similar size produces
the ringing but not the thump.

Basketballs (and several other types of ball, including volleyballs,
[American] footballs and soccer [football outside the U.S.] balls) are
inflated with air to an overpressure of several psi.  This overpressure
gives them their stiffness and resilience.  We describe the outer skin of
the ball (which is made of various combinations of rubber, leather, and
nylon fiber) as an inextensible but perfectly flexible membrane.
Inextensibility is a fair approximation, as shown by the fact that these
balls do not much increase dimension upon inflation.  Perfect flexibility
is, at best, a rough approximation, but is necessary to describe the
mechanics of bounce without resorting to a numerical elastodynamic treatment
of the skin.  In addition, we assume normal incidence and no rotation.

\section{Energetics}

\begin{figure}
\centerline{\includegraphics{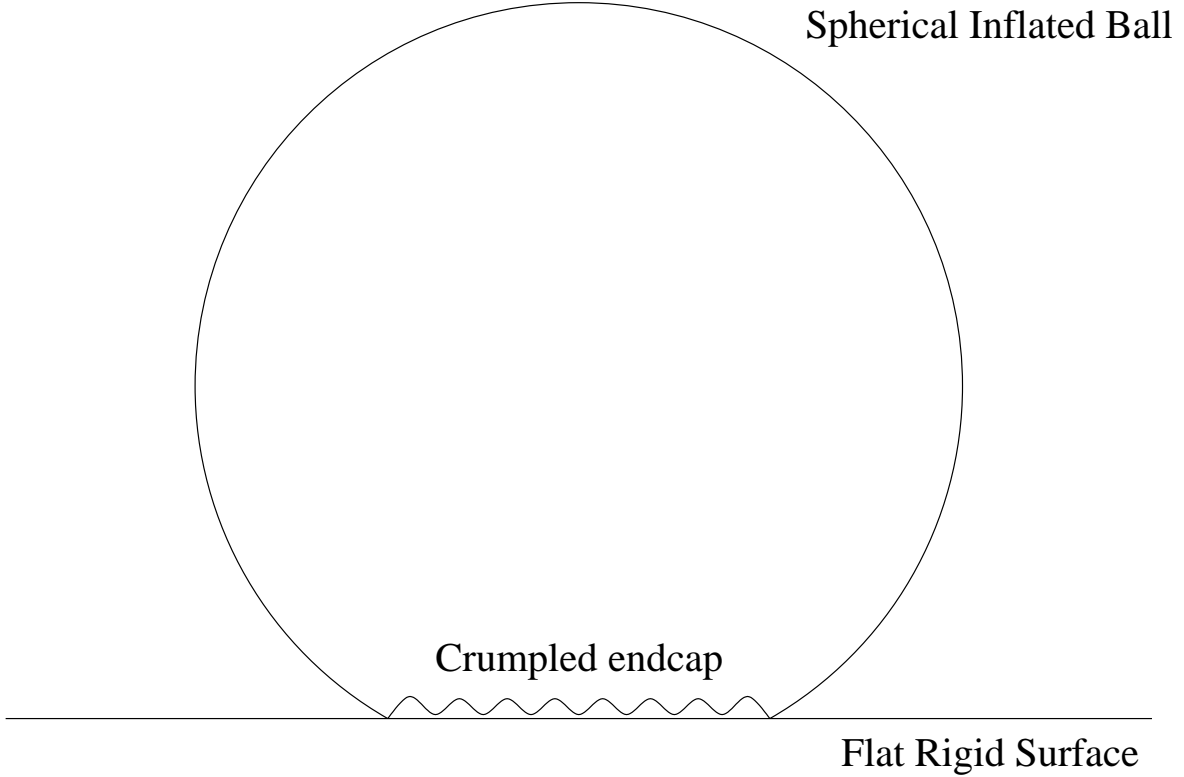}}
\caption{\label{thumpfig} An inflated (pressurized) spherical ball made of
an inextensible but flexible skin recoiling from a stiff flat surface.  The
ball retains its spherical shape except where in contact with the surface.
The skin of the flattened spherical cap crumples to accommodate its reduced
surface area.}
\end{figure}

When such a ball strikes a stiff immovable flat surface the portion of the
ball pressing against the surface loses its spherical shape.  The remainder
of the ball remains spherical because inextensibility implies that latitude
lines cannot stretch, and the volume is maximized if they retain their
pre-impact spherical geometry rather than telescoping.  Maximization of
volume also implies that the contacting portion of the ball, formerly a
spherical cap, is pressed flat against the rigid surface (Figure
\ref{thumpfig}).  This requires a complex pattern of crumpling, which we do
not calculate.  Because the skins of real balls have a finite thickness
(about 3.5 mm for basketballs), the approximation of an infinitely flexible
membrane is not accurate, but we make this approximation to permit a simple
analytic treatment.

Flattening the end cap of a spherical ball, as shown in Figure
\ref{thumpfig}, requires a mechanical work
\begin{equation}
W = \int P\,dV \approx P \Delta V = \pi P (ax^2 - x^3/3) \approx \pi P a
x^2,
\end{equation}
where $x$ is the depth of flattening of the ball (the distance its center
has traveled normal to the surface after first contact) and $a$ is its
unflattened radius.  Taking $x \ll a$, we have approximated $P$, the excess
pressure inside the ball over ambient pressure ($P$ is also known as the
gauge pressure), as a constant.

The mechanical work has the form of a simple harmonic oscillator potential
\begin{equation}
W = {1 \over 2} k x^2
\end{equation}
with spring constant
\begin{equation}
k = 2 \pi P a.
\end{equation}
As a result, the motion of the ball when it is in contact with the surface
is simple harmonic, with
\begin{equation}
x = x_0 \sin{\omega t} \label{motion}
\end{equation}
for $0 \le t \le \pi/\omega$, with frequency
\begin{equation}
\label{freq}
\omega = \left({2 \pi P a \over M}\right)^{1/2},
\end{equation}
where the mass $M$ includes the mass of the skin of the ball, its contained
air, and the induced mass of the surrounding airflow.  To an excellent
approximation for most balls used in sports, $M$ is given by the mass of the
solid (actually, polymer) skin.

The ball remains in contact with the surface for a half-cycle of this
oscillation, independent of its velocity $v_0 = x_0 \omega$ at impact.  If
dropped from a height $h$, then $v_0 = \sqrt{2gh}$, ignoring air drag during
its fall (for $h = 1$ m air drag introduces an error $\sim 10$\%).

\section{Monopole emission---the ``thump''}

The changing volume of the ball produces a monopole source\cite{morse48} of
sound described by the pressure field
\begin{equation}
p_{rad}(r,t) = {\rho \over 2 \pi r} Q^\prime(t-r/c),
\end{equation}
where $\rho$ is the density and $c$ the sound speed of the surrounding
fluid, the volume flow rate $Q(t) = {dV \over dt}$, and we have multiplied
the standard result by a factor of 2 to allow for the fact that at the
surface of an infinite rigid slab sound is radiated into only $2\pi$ sterad.
Using $V \approx V_0 - \pi a x^2$ and Eq.~(\ref{motion}) we obtain
\begin{equation}
p_{rad}(r,t) \approx - {\rho a v_0^2 \over r}\cos{2 \omega (t - r/c)},
\label{prad}
\end{equation}
for $0 \le t - r/c \le \pi/\omega$, and zero otherwise.

\begin{figure}
\includegraphics{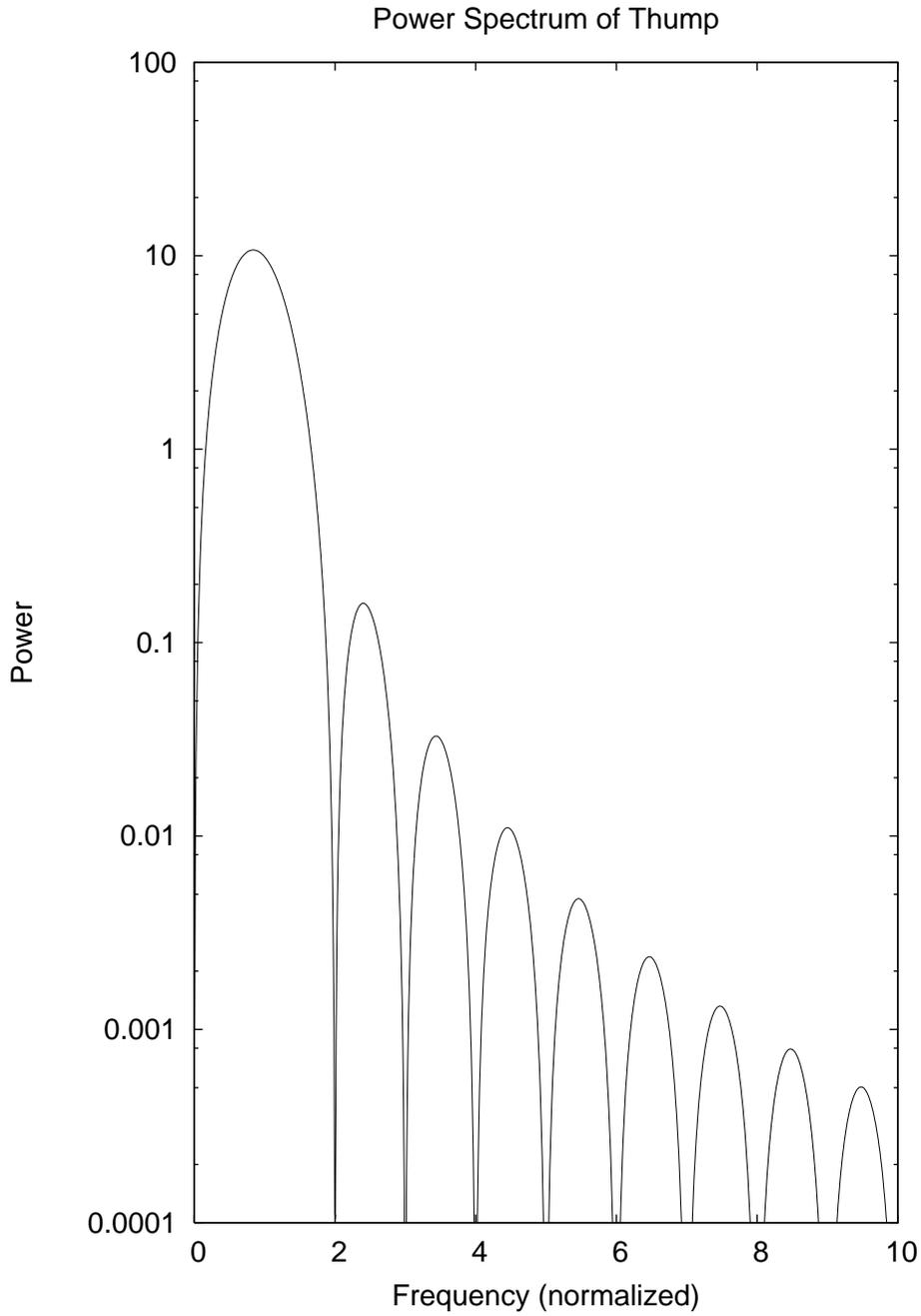}
\caption{\label{thumpspect} Power spectrum of the ``thump''.  The frequency
is normalized to that given by Eq.~\ref{freq}.}
\end{figure}

Standard basketballs have $M \approx 600$ g, $P \approx 5.5 \times 10^5$
dynes/cm$^2$ (8 psi), and an internal radius $a \approx 11.4$ cm, yielding
$\omega \approx 256$ s$^{-1}$.  The acoustic pulse described by
Eq.~(\ref{prad}) varies sinusoidally at a frequency $2 \omega$,
corresponding to 82 Hz, but because only one full cycle of this frequency is
present its Fourier transform (Figure \ref{thumpspect}) is broad, giving a
dull ``thump''.  Taking
$v_0 = 443$ cm s$^{-1}$ ($h = 100$ cm) and $\rho = 1.19 \times 10^{-3}$ gm
cm$^{-3}$ (dry air at 20$^\circ$ C and standard pressure) yields a peak
pressure amplitude of 8.9 dynes cm$^{-2}$, or 93 dB (referred to the
standard 0 dB level of $2 \times 10^{-4}$ dyne cm$^{-2}$) at a range $r =
300$ cm.  This explains the surprising loudness of the ``thump''.

\section{Dipole emission---the ``ring''}

In addition to the ``thump'', a high-pitched ringing sound is also heard.
When a softer rubber ball, inflated to low pressure, such as often used as a
child's toy, is used, the ``thump'' is less audible because $\omega$ is very
low, and the ringing is more striking.  The frequency is that of the lowest
eigenmode of the oscillation of the air inside the ball.

In the inextensible approximation the skin of the ball may be considered to
be a rigid spherical shell after contact with the flat surface is broken.
The governing equation for sound waves is 
\begin{equation}
\nabla^2 p + {\omega^2 \over c^2}p = 0,
\end{equation}
where $\omega$ is now the frequency of the acoustic wave.  Azimuthally
symmetric modes are given (inside the ball) by
\begin{equation}
p(r,\theta) = P_\ell(\cos\theta) j_{\ell}(kr),
\end{equation}
where the wave vector $k \equiv \omega/c$.  $P_\ell$ is the Legendre
polynomial of order $\ell$ and $j_\ell$ is the spherical Bessel function of
that order.

The value of $k$ is determined by the boundary condition $dj_\ell(kr)/dr =
0$ at $r = a$ because at the outer surface $0 = v_r = {\partial p \over
\partial r}\, P_\ell(\cos\theta)/(i\omega\rho)$.  The boundary condition on
the parallel component of velocity is not applicable outside a boundary
layer of thickness ${\cal O}\sqrt{\nu/\omega_1} \sim  0.05$ mm, where $\nu$
is the kinematic viscosity.

The lowest frequency mode
is found for $\ell = 1$, corresponding to a mode in which all the air moves
in the same direction at any one time, and the diameter is approximately a
half-wavelength ($ka = 2.082$).  The $\ell = 0$ mode has a velocity node at
the origin as well as at the surface, so the diameter is approximately a
full wavelength ($ka = 4.494$).  Its frequency is roughly twice as high as
that of the $\ell = 1$ mode.  Hence the lowest mode frequency is
\begin{equation}
\omega_1 = {2.082 c \over a}.
\end{equation}
At a temperature of 20$^\circ$ C ($c = 3.43 \times 10^4$ cm s$^{-1}$) we
find $\omega_1 = 6.26 \times 10^3$ s$^{-1}$ (997 Hz).

It is possible to estimate the amplitude of this mode.  It is excited, in
the inextensible approximation, by the force the rigid surface exerts on the
air when the ball is in contact with the surface. In the inextensible
membrane approximation this force is transmitted through the flattened
portion of the skin to the contained air.  A slowly-varying (at the
frequency $\omega$) pressure gradient in that air accelerates the skin
(which contains nearly all the mass).  

The momentum imparted to the oscillating air then equals the Fourier
transform of the applied force $F(t)$ at the mode frequency $\omega_1$.  The
resulting momentum is 
\begin{equation}
P_{\omega_1} = \int^{\pi/\omega}_0 F(t) \exp{i\omega_1 t}\,dt =
\int^{\pi/\omega}_0 p A(t) \exp{i\omega_1 t}\,dt,
\end{equation}
where $A(t)$ is the area of contact.  Using $A(t) = 2 \pi a x$ and 
Equation~(\ref{motion}) we find
\begin{equation}
P_{\omega_1} = 2 \pi a p x_0 \int^{\pi/\omega}_0 \sin{\omega t} \exp{i
\omega_1 t}\,dt.
\end{equation}
The integral is complex, but we are only interested in its modulus
\begin{equation}
\vert P_{\omega_1}\vert = 2 \pi a p x_0 {\omega \over \omega_1^2 - \omega^2}
\left[2\left(1 + \cos{\pi \omega_1/\omega}\right)\right]^{1/2}.
\end{equation}
Because $\omega_1 \gg \omega$ the factor in the square bracket is
unknowable, but its root mean squared value may be found by averaging over a
full cycle of the cosine.  The resultant root mean squared momentum
\begin{equation}
\langle P_{\omega_1} \rangle = 2 \sqrt{2} \pi a p x_0 {\omega \over
\omega_1^2 - \omega^2}.
\end{equation}

It is possible to integrate the velocity field over the interior of the ball
to find the pressure and velocity amplitude corresponding to $P_{\omega_1}$.
Writing $p_1(r,\theta) = a_1 P_1(\cos\theta) j_1(kr)$ we find
\begin{equation}
a_1 = {p a x_0 \over c^2} {\omega^4 \over \omega_1^2 - \omega^2} {\sqrt{2
(1 + \cos{\pi \omega_1/\omega})} \over 3.463}.
\end{equation}

We are more interested in the acoustic radiation by this mode, which can be
obtained directly from $P_{\omega_1}$.  The oscillation of the air within
the ball produces an opposite oscillation of the ball, with velocity
amplitude $U_1 = P_{\omega_1}/M$.  The dipole approximation is not valid
because $ka = 2.082$ rather than $ka \ll 1$.  However, Morse\cite{morse48}
tabulates numerical solutions for general $ka$.  The resulting radiation
field
\begin{equation}
p_{rad}(r,t) = {\rho c U_1 \over k r D_1} \cos\theta \cos{\omega(t - r/c)},
\end{equation}
where for $ka=2.082$ the factor $D_1 = 0.531$.  For our previous parameters
$x_0 = v_0/\omega = 1.70$ cm and we find $\langle P_{\omega_1} \rangle =
750$ gm cm s$^{-1}$, $U_1 = 1.25$ cm s$^{-1}$ and the amplitude of $p_{rad}$
at $r = 300$ cm and $\theta = 0$ is 1.87 dyne cm$^{-2}$, or 79 dB.  This
pressure amplitude is 14 dB below that of the low frequency ``thump'', but
is readily audible.

\section{The inextensible approximation}

The validity of the inextensible membrane approximation may be quantified by
comparing the frequency $\omega$ which describes the bounce to the
frequency $\omega_{el}$ of the ``breathing'' mode in which the skin
oscillates, spherically symmetrically, about its equilibrium radius.  The
inextensible approximation corresponds to the limit $\omega_{el}/\omega \to
\infty$.

By force balance we see that inflating the ball to an overpressure $P$
increases its radius by an amount
\begin{equation}
\delta a = {P a^2 \over h E},
\end{equation}
where $h$ is the skin's thickness and $E$ its Young's modulus.  Young's
modulii of rubber are found\cite{ca94} over a wide range, roughly
$10^7$--$10^9$ dyne cm$^{-2}$, while that of leather is typically\cite{ict29}
about $5 \times 10^8$ dyne cm$^{-2}$.  Basketballs generally are stiffened
with wound nylon filament with a modulus\cite{ca94} of about $3 \times
10^{10}$ dyne cm$^{-2}$.  If $P$ is measured with a pressure gauge, then
measuring $\delta a$ may be the most convenient means of determining $Eh$.

The breathing mode oscillation is described by a potential energy
\begin{equation}
U = -P\,dV + 2 \times 4 \pi a^2 h {u \sigma \over 2},
\end{equation}
where $u$ is the strain and $\sigma$ the stress and the first factor of two
comes from the fact that the skin is stretched along two orthogonal axes.
For small strains $u \approx \delta a /a$, $\sigma \approx E \delta a/a$ and
\begin{equation}
U \approx -4 \pi a^2 P \delta a + 4 \pi (\delta a)^2 h E,
\end{equation}
yielding an effective spring constant $k_{breathe} = \partial^2 U/\partial
\delta a^2 = 8 \pi h E$.  Then, using $M = 4 \pi a^2 h \rho_{skin}$,
\begin{equation}
\omega_{el} = \left({2 E \over \rho_{skin} a^2}\right)^{1/2}.
\end{equation}
The static inflation pressure drops out because the elastic response is
assumed to be linear.

The ratio of the breathing mode frequency to the bounce frequency is
\begin{equation}
{\omega_{el} \over \omega} = \left({4 E h \over P a}\right)^{1/2} = \left({
4 a \over \delta a}\right)^{1/2}.
\end{equation}
For an inextensible membrane $E \to \infty$ and $\omega_{el}/\omega \to
\infty$.  The ratio $a/\delta a$ is readily measured as the ball is inflated.
A rough measurement on a Spalding ``NBA Indoor/Outdoor'' basketball yielded
$\delta a/a = 0.009$ for $P = 8$ psi, the breathing mode frequency
$\omega_{el}/2\pi$ is about 900 Hz (accidentally close to the ringing
frequency), $Eh = 7 \times 10^8$ erg cm$^{-2}$ and $E = 2 \times 10^9$ erg
cm$^{-3}$.  The stiffness is largely provided by the nylon filaments.

The ratio $\omega_{el}/\omega = 21$ is substantially, but not enormously
larger than unity, so inextensibility should be a reasonable approximation.
The infinite flexibility approximation may not be accurate, but is
not so readily quantified, and the actual dynamics at impact involves such
complications as friction with the surface.

The most important consequence of the breakdown of the inextensibility
approximation is that impact excites the breathing mode.  After the ball
breaks contact with the surface it continues to radiate.  The rubbery
materials of which basketballs are made have large loss coefficients, so
this oscillation and radiation damp in about a single cycle.

In the opposite (extensible membrane, incompressible filling fluid)
approximation the volume is not changed at impact and there is neither
``thump'' nor ``ring''.  This limit holds accurately for a water-filled
latex balloon.  An air-filled latex balloon is an intermediate case, and
some ``thump'' is heard.  There is no ringing because the stretchable and
low mass membrane permits internal modes to damp rapidly by radiation.

\begin{acknowledgments}
I thank the Los Alamos National Laboratory, where this project was conceived
during the lunch break of a committee meeting, for hospitality, and my son
Alexander Z. Katz for bouncing a basketball on the floor (terrazzo over 8
inches of concrete) of our house.
\end{acknowledgments}

\bibliography{thump.bib}
\end{document}